\documentclass[preprintnumbers,prd,nofootinbib]{revtex4-1}

\usepackage{amssymb,amsmath}
\usepackage{graphicx,color}
\usepackage{amscd}

\newcommand{\m}{ \mathcal{M} }
\newcommand{\simgt}{\lower.5ex\hbox{$\; \buildrel > \over \sim \;$}}
\newcommand{\simlt}{\lower.5ex\hbox{$\; \buildrel < \over \sim \;$}}

\begin{document}
\title{
Random matrix model for chiral and color-flavor locking condensates
} 

\author{T.~Sano$^{a,b}$ and K.~Yamazaki$^{a,b}$}
\address{%
$^a$Institute of Physics, The University of Tokyo, \\
Tokyo 153-8902, Japan\\ 
$^b$Department of Physics, The University of Tokyo, \\
Tokyo 113-0033, Japan\\ 
}

\date{\today}

\begin{abstract}
We study the phase diagram of a chiral random matrix model 
with three quark flavors at finite temperature and chemical potential, 
taking the chiral and diquark condensates 
as independent order parameters.
Fixing the ratio of the coupling strengths 
in the quark-antiquark and quark-quark channels 
applying the Fierz transformation, 
we find that the color-flavor locked (CFL) phase 
is realized at large chemical potential, 
while the ordinary chirally-broken phase appears 
in the region with small chemical potential.
We investigate responses of the phases
by changing small quark masses in the cases 
with three equal-mass flavors and with 2+1 flavors. 
In the case with three equal-mass flavors, 
we find that the finite masses make
the CFL phase transition line move 
to the higher density region.
In the case with 2+1 flavors,
we find the two-flavor color superconducting phase 
at the medium density region
as a result of the finite asymmetry between the flavors,
as well as the CFL phase at higher density region.
%
%
\end{abstract}

\maketitle

\section{Introduction}

Mapping of QCD phase diagram at finite temperature and 
density\cite{CMP_QCD, Fukushima:2010bq, Alford:2007xm, Buballa:2003qv, Hatsuda:1994pi} 
is one of the most challenging issues in the theoretical and experimental physics
and is significant to 
the heavy-ion collision experiments 
and the structures of the neutron stars.

At finite temperature and zero or small baryon density, 
a number of 
investigations on the QCD phase transition 
are made both with the lattice QCD simulations\cite{DeTar:2011nm} 
and with the model calculations\cite{Hatsuda:1994pi}. 
Lattice QCD simulations suggest that the phase transition becomes 
smooth crossover 
in the realistic case with two light and one heavier quark flavors,
and many model calculations are consistent 
with this result. 
%
%
At finite density, however, the situation is
uncertain because lattice QCD simulations are
still challenging 
at finite chemical potentials
with low temperatures\cite{deForcrand:2010ys}.
In such a region, it is important 
to employ models for qualitative and quantitative calculations 
on the phase diagram. 
Naively, a large baryon number density may cause overlaps of baryons, 
which invalidates a concept of confined colors in a baryon, 
resulting in deconfinement, which may be followed 
by chiral phase transition.

Furthermore, at asymptotically large density, 
the appearance of a color-superconducting (CSC) phase is 
also expected, 
where a weak coupling theory is applied 
and the Cooper instabiltiy of the Ferimi sphere is 
inevitable\cite{Barrois:1977xd,Bailin:1983bm}. 
The color superconducting phases are characterized 
by the diquark condensates. 
One-gluon exchange interactions tell us 
that the color-antitriplet channel is attractive.
Therefore, with the Pauli principle, 
condensates in the color- and flavor-antitriplet and spin antisymmetric 
channel is expected
%
.
Such condensates can be expressed as
\begin{align}
s_{AA'}
=
\left<
\bar{\psi^c} C \gamma_5 \tau_A \lambda_{A'} \psi
\right>
,
\end{align}
where $\psi$ is the quark field, and
$\bar{\psi^c}=\psi^T C$ with $C$ the charge conjugation operator. 
$\tau_A$ and $\lambda_A$, where $A=2, 5$ and $7$, 
are the antisymmetric generators of 
SU($N_f$) flavor and SU($N_c$) color groups, respectively.

One of the most striking features in the CSC phases 
is the formation of the color-flavor locking (CFL) condensate.
At sufficiently high density, the finite current quark masses 
for up, down and strange quark flavor are neglected so that 
the system can be treated as the chiral limit.
In the CFL phase, which is characterized by
$s_{AA'}=\Delta \delta_{AA'}$ with nonzero $\Delta$,
SU(3)$_L \otimes$ SU(3)$_R \otimes$ SU(3)$_c$ symmetry of the system 
breaks down to its subgroup of SU(3)$_{L+R+c}$.
This breaking pattern 
is possible owing to the miraculous matching of 
the (effective) number of flavors $N_f=3$ and the number of colors $N_c=3$.

To investigate the QCD phase diagram 
at finite temperature and chemical potential, 
chiral random matrix (ChRM) models provide us 
of a qualitative way 
from a viewpoint of the symmetry\cite{ChRM, Halasz:1998qr}.
In a conventional ChRM model, 
the Dirac operator is set as a random matrix, 
which has the same symmetry with QCD, 
and the partition function is defined as 
the average of the determinant of the Dirac operator 
over the matrix elements with the Gaussian distribution. 
The random distribution of the matrix elements mimics the 
complex dynamics of the gluon fields. 
The Gaussian model can be solved exactly in the thermodynamic limit. 
Although the model is constructed in such a simple way, 
the resulting phase diagram has a rich structure.
In the chiral limit, 
the phase transition becomes second-order in 
the small chemical potential $\mu$ region, 
while it becomes first-order in the large $\mu$ region. 
First- and second-order phase transition lines are connected at 
the tricritical point (TCP).
This result is consistent with NJL models\cite{Asakawa:1989bq, Barducci:1989wi}.

The extension of the ChRM model to the case with the CSC phase 
has already been studied
by Vanderheyden and Jackson
\cite{Vanderheyden:1999xp}\footnote{
Also, there are studies on the diquark condensates with $N_c=2$
using ChRM models\cite{Vanderheyden:2001gx,Klein:2004hv}.
}.
In their study, 
the Dirac matrix is extended to have the indices of color and spin explicitly 
with the real random matrices 
corresponding to real gluon fields in QCD. 
After the integration over the random matrices, 
the model produces the quark-quark interaction terms, 
which are responsible for the diquark condensates,  
as well as quark-antiquark interaction terms 
responsible for the chiral condensates. 
The resulting phase diagram has the diquark condensed phase 
at large $\mu$ region, 
while the chirally-broken phase 
at small $\mu$ region, 
if the ratio of the quark-antiquark and the quark-quark coupling is taken 
so that the Dirac operator of the model 
has the same symmetry with that of QCD.
Note that because the model 
in Ref.~\cite{Vanderheyden:1999xp} 
contains two quark flavors, 
the CSC phase is the
two-flavor color superconducting (2SC) phase, 
where only two of three colors of fermions 
participate in the diquark pairing.

It is then natural to ask whether 
it is possible to extend the ChRM model to the case with three flavors, 
and whether the CFL phase can appear as the ground state in a high density region.
We answer ``yes'' to this question 
by constructing the ChRM model containing 
three flavors and colors, 
and show a phase diagram with the chirally-broken 
and the CFL phase. 
As a simple application of this model, 
we also focus on the response of the model by changing the quark masses. 
By setting the strange quark mass different 
from the other two quark flavors, 
we observe that the 2SC
phase appears on the phase diagram 
at the moderate values of the chemical potential,
as well as the CFL phase in the larger chemical potential region.

This paper is organized as follows. 
We introduce an extended ChRM model 
with chiral and CFL condensations in sec.~2, 
and derive its effective potential in sec.~3.
The model phase diagrams are presented 
and discussed in the case with 
three equal-mass flavors and 2+1 flavors in sec.~4 and 5, respectively. 
Sec.~6 is devoted to a summary and discussion.

\section{Random matrix model with chiral and diquark condensations}

In this section, we introduce a chiral random matrix model which mimics 
QCD partition function with three quark flavors, 
extending the two-flavor case in Ref.~\cite{Vanderheyden:1999xp}.
We denote three quark masses by $m_f$ with $f=$u, d and s.




Keeping in mind the (extended) Banks-Casher relations, 
which relate Dirac soft modes 
not only with the chiral condensates\cite{Banks:1979yr}, 
but also with the diquark condensates\cite{Fukushima:2008su}, 
we consider the truncated Dirac matrix $D$ in 
low-lying quark excitation, or zero-mode space.
We assume $D$ can be separated as $D=R+C$, where $R$ is 
a random part, which represents the complex gluon dynamics, 
and $C$ the non-random, deterministic part, 
which is responsible for the matter effects.

For simplicity, we first set the matter effects turned off, i.e. $C=0$, 
and focus on the random matrix $R$.
In this case, the truncated Dirac matrix should have 
the symmetries of the Dirac operator in the vacuum,
the chiral symmetry, $\{\gamma_5, R\}=0$, 
and the anti-hermiticity, $R^\dagger = -R$. 
Within these restrictions, 
the Dirac matrix generally has 
nonzero matrix elements 
only in the off-diagonal blocks
\begin{align}
R=
\left(
\begin{matrix}
0	&	iW\\
iW^\dagger	&	0
\end{matrix}
\right)
,
\label{e:diracop}
\end{align}
in the chiral representation, 
$\gamma_5 = {\rm diag}(+1,-1)$, 
where $W$ is a complex matrix%
\footnote{%
One can take $W$ generally to be 
an $N_+ \times N_-$ rectangular matrix. 
In this case, the Dirac matrix has 
$|\nu |=|N_+ - N_-|$ exact zero eigenvalues, 
which represents the index theorem
with the background gauge field having the topological charge $\nu$.
Exploiting this fact, we can introduce the effect 
of the axial anomaly in the ChRM models
\cite{Janik:1996nw,Sano:2009wd}.
In this study, however, we always take $W$ to be square 
and the anomaly effect is neglected. See the discussion in sec.~6.
}%
.
In the conventional ChRM models\cite{ChRM}, 
$W$ is taken to be a general complex matrix whose elements 
are independently distributed according to the Gaussian distribution. 

To consider the diquark condensations, however, 
it is crucial to treat the color and spin 
indices explicitly.
Following the construction in Ref.~\cite{Vanderheyden:1999xp}, 
we express $W$ as a direct product of the spin, 
color and zero-mode matrices, whose total dimension is
$2 \times N_c \times N$, 
where 2 is the size of the spin space, 
$N_c$ the color space, and $N$ the zero-mode space.
We adopt the form of $W$ as
\begin{align}
W
=
A^{\nu a} 
(\sigma_\nu \otimes \lambda_a)
\label{e:w}
,
\end{align}
where 
$\sigma_\nu = (1, -i \sigma_i)$ with $\sigma_i$ the Pauli matrix, 
$\lambda_a$ is a generator of SU($N_c$), 
and $A^{\nu a}$ is a $N\times N$ random matrix.
Since the random matrix $A^{\nu a}$ corresponds to the gauge field in QCD, 
we choose $A^{\nu a}
$ to be a real matrix. 

The matter effects are introduced 
as the non-random external fields in the Dirac operator.
A simple way\cite{Halasz:1998qr} is to add a constant matrix $C$ to 
the random matrix (\ref{e:diracop}) with
\begin{align}
C
=
\left(
\begin{matrix}
0 & \omega\\
\omega & 0
\end{matrix}
\right)
=
\omega \otimes \gamma_0
,
\label{e:mem}
\end{align}
where the $2 \times N_c \times N$-dimensional matrix $\omega$ is defined as
\begin{align}
\omega
=
\left(
\begin{matrix}
(\mu +i T)\mathbf{1}_{N/2}\otimes \mathbf{1}_{\rm spin}\otimes \mathbf{1}_{N_c}  & 0\\
0 & (\mu -i T)\mathbf{1}_{N/2}\otimes \mathbf{1}_{\rm spin}\otimes \mathbf{1}_{N_c} 
\end{matrix}
\right)
.
\end{align}
$T$ and $\mu$ are an effective temperature and quark chemical potential, 
respectively. 
$\mathbf{1}_{\rm spin}$ denotes $2\times 2$ identity matrix 
in the spin space. 
A total Dirac matrix, $D=R+C$, also has chiral symmetry, 
$\{D, \gamma_5 \}=0$, but not anti-hermiticity,
$D^\dagger \neq - D$ if $\mu \neq 0$.
Two relative signs between $T$ and $\mu$
corresponds to the two lowest Matsubara frequency, $\pm \pi T$. 
The inclusion of two signs 
reproduces the invariance of the partition function
under the charge conjugation transformation, $\mu \to -\mu$.

Using the Dirac matrix $D$, 
the ChRM model partition function is defined as
\begin{align}
Z
=
\int [dA]
\prod_{f=1}^{N_f}
\det (D+m_f)
e^{-2N \Sigma^2 \sum_{a,\nu, i, j} (A_{ij}^{\nu a})^2}
,
\label{e:pf}
\end{align}
where the integral is defined over real elements of 
random matrices $A^{\nu a}$ with the Gaussian weight. 
The parameter $\Sigma$, which fixes variance of the Gaussian distribution, 
gives a scale to the model.
Generally, $\Sigma$ may change for each $a$ and $\nu$, 
but by ensuring the color and Loerntz symmetry, 
their values should be equal.

Before solving the model, we make two remarks on 
the treatment of the ChRM model 
comparing to that in Ref.~\cite{Vanderheyden:1999xp}. 
First, in Ref.~\cite{Vanderheyden:1999xp}, 
the authors examine not only the form of the random Dirac operator 
(\ref{e:diracop}), but also the case 
where the Dirac operator breaks the symmetry 
which QCD Lagrangian holds.
In such general cases, the random matrix $R$ is taken as
\begin{align}
(R)_{\alpha \beta}
=
\sum_{C a}
X^{C a}_{\alpha \beta}(\Gamma_C)_{\alpha \beta} \otimes \lambda_a
,
\end{align}
where the chiral indices $\alpha, \beta =$ R, L, 
$X^{C a}_{\alpha \beta}$ is a random matrix, 
and $\Gamma_C$ is the independent gamma matrix in four dimension, 
$C = 1, \dots 16$.
A set of $\Gamma_c$ can be separated into the subsets forming
a Lorentz scalar, pseudo-scalar, vector, axial-vector, and tensor.
In Eq.(\ref{e:diracop}) and (\ref{e:w}),
we have chosen $X^{C a}_{\alpha \beta}$ to be nonzero
only for the vector content of the gamma matrices.
This is a natural choice because we consider that
the random matrices model the gluon fields, which form a Lorentz vector.
Indeed, 
nonzero components for the scalar, pseudo-scalar 
and tensor break the chiral symmetry explicitly,
and that for the axial-vector does the antihermiticity.
If such non QCD-like random matrices are allowed, 
one can vary the ratio of 
the coefficients of the quark-antiquark and the quark-quark interaction 
channels, which is denoted as $B/A$ in the next section.
The evolution of the phase diagram with $B/A$ changed
was completely investigated in Ref.~\cite{Vanderheyden:1999xp}
for the case with two flavors.
In our study, however, we only focus on the case in which 
the model has the same type of the interaction with QCD, 
because the condition fixes the topology of the phase diagram 
without ambiguity and 
we can solely concentrate on the response of the phase diagram 
by changing the quark masses.

Second, the scheme of the temperature dependence is different 
from that in Ref.~\cite{Vanderheyden:1999xp}. 
We use the matter effect matrix proportional to the identity in the flavor space, 
while, in Ref.~\cite{Vanderheyden:1999xp}, 
the sign of the temperature $T$ is antisymmetric in the space of two flavors. 
As a result of this difference,
our model can not be reduced to the model in Ref.~\cite{Vanderheyden:1999xp}
in the two flavor limit. 
Further discussion will be given 
by comparing the effective potentials 
in the later section.

Finally, note that the partition function (\ref{e:pf})
has 
SU($N_C$)$_c \otimes$SU($N_f$)$_L \otimes$SU($N_f$)$_L$ 
global symmetry, but not local gauge symmetry.
To be exact, it is then appropriate to call the diquark condensed phase 
a BEC state, not a BCS state.
In the remaining part of this paper, 
we use the word, ``diquark condensates'' to indicate such condensates,
but discuss them comparing to the BCS states expected in QCD 
at finite density.



\section{The effective potential}

In this section, we derive the effective potential of the ChRM model 
defined in Eq.(\ref{e:pf}). 
The derivation is almost parallel 
to that in Ref.\cite{Vanderheyden:1999xp}. 
We present the derivation in three steps, 
and then make a few remarks.

\subsection{Gaussian integral} 
The first step is to integrate out the Gaussian integral variables $A_{ij}^{a\mu}$. 
For this purpose, we first express the determinant 
in the partition function (\ref{e:pf}) 
in the form with the fermion integrals: 
\begin{widetext}
\begin{align}
\prod_{f=1}^{N_f}
\det (D + m_f)
&=
\int [d \psi^\dagger ][d \psi ]
\exp \left[ - \sum_f \bar \psi^f (D +m_f)\psi^f \right]
\nonumber
\\
&=
\int [d \psi^\dagger ][d \psi ]
\exp 
\left[ 
- i 
J^{ij}_{a\nu} A^{a\nu}_{ij}
-
\sum_f
\bar \psi^f (C + m_f) \psi^f 
\right]
,
\end{align}
\end{widetext}
where $\psi^f = (\psi_R^f, \psi_L^f)^T$ and 
$\bar \psi^f = (\psi_L^{f\dagger}, \psi_R^{f\dagger})$ 
are $4 \times N_c \times N$ Grassmann vectors, and
\begin{align}
J_{a\nu}^{ij}
=
\sum_f
\left(
\psi_{Li}^{f\dagger}     \sigma_\nu \lambda_a \psi_{Lj}^f
+
\psi_{Rj}^{f\dagger}\sigma_\nu^\dagger \lambda_a \psi_{Ri}^f
\right)
\end{align}
is a fermion bilinear.

Applying the Gussian integral formula
up to the constant,
$
\int dx e^{-\alpha x^2 +\beta x}
=\exp (\beta^2/(4 \alpha))
$
, 
to the $A_{a\nu}^{ij}$ integral separately for each indices 
$i, j, a$ and $\nu$, 
we obtain analytically
\begin{align}
\int [dA]
e^{-i J_{a\nu}^{ij}A^{a\nu}_{ij}}
e^{-2N \Sigma^2 (A^{a\nu}_{ij})^2}
=
\exp \left[ -\frac{1}{8 N \Sigma^2} (J_{a\nu}^{ij})^2\right]
\label{e:fourpoint}
,
\end{align}
where the summations over $i,j,a$ and $\nu$ 
should be understood on the right-hand side.

\subsection{Fierz transformation}
In this step, we expand the square of the fermion bilinear 
$(J_{a\nu}^{ij})^2$ and realign the four point vertices, 
by applying the Fierz transformation formula. 
%
%
The square of $J_{a\nu}^{ij}$ is expanded as
\begin{align}
(J_{a\nu}^{ij})^2
=&
2 
\psi_{R i}^{f \dagger}\sigma_\nu^\dagger \lambda_a \psi_{R j}^f
\psi_{L j}^{g \dagger}\sigma_\nu        \lambda_a \psi_{L i}^g
\nonumber
\\
&+
\psi_{L i}^{f \dagger}\sigma_\nu        \lambda_a \psi_{L j}^f
\psi_{L i}^{g \dagger}\sigma_\nu        \lambda_a \psi_{L j}^g
+
\psi_{R i}^{f \dagger}\sigma_\nu^\dagger \lambda_a \psi_{R j}^f
\psi_{R i}^{g \dagger}\sigma_\nu^\dagger \lambda_a \psi_{R j}^g
.
\end{align}
The first term represents the quark-antiquark interactions, 
and the other terms the quark-quark interactions. 
The former is responsible for the formation of the chiral condensates 
and the latter for the diquark condensates.

Using Fierz transformations, these four-fermion terms 
are rearranged so that 
in each fermion bilinear terms, 
zero-mode indices $i$ and $j$ are contracted.
%
%
At this point, we assume that the chiral condensates are formed 
only in the color-singlet, scalar channels, 
and that the diquark condensates are formed only in the 
spin-antisymmetric, flavor- and color-antitriplet, scalar channels. 
With these assumptions, relevant interaction terms are drastically 
reduced and shown explicitly as%
\footnote{%
For the Fierz transformation formulae, 
see, for example, Ref.~\cite{Buballa:2003qv}.
} 
\begin{align}
\psi_{R i}^{f \dagger}\sigma_\mu^\dagger \lambda_a \psi_{R j}^f
\psi_{L j}^{g \dagger}\sigma_\mu        \lambda_a \psi_{L i}^g
=
-
\frac{2(N_c^2 - 1)}{N_c^2}
\psi_{R}^{f \dagger} \psi_{L}^g
\psi_{L}^{g \dagger} \psi_{R}^f
+\dots
\end{align}
for quark-antiquark channels and
\begin{align}
\psi_{L i}^{f \dagger}     \sigma_\mu \lambda_a \psi_{L j}^f
\psi_{L i}^{g \dagger}     \sigma_\mu \lambda_a \psi_{L j}^g
=
-
\frac{N_c + 1}{2 N_c}
\psi_L^\dagger     \tau_A \lambda_{A'} \psi_L^c
\psi_L^{c \dagger} \tau_A \lambda_{A'} \psi_L
+\dots
\end{align}
for quark-quark channels, with the same except L $\to$ R,
where the charge conjugation fields are defined as
$\psi^c \equiv C \bar \psi^T = (\psi^c_L, \psi^c_R)^T$ 
and
$
\bar \psi^c \equiv \psi^T C 
= (\psi^{c \dagger}_R, \psi^{c \dagger}_L)
$.
The dots denote the terms irrelevant to 
the formation of the condensates we focus.


Neglecting the irrelevant terms,
we finally obtain the simple form of four-point interaction as
\begin{align}
(J_{a\nu}^{ij})^2
=&
-
2
G_\chi
\psi_{R}^{f \dagger} \psi_{L}^g
\psi_{L}^{g \dagger} \psi_{R}^f
\nonumber
\\
&-
G_\Delta
\psi_L^\dagger     \tau_A \lambda_{A'} \psi_L^c
\psi_L^{c \dagger} \tau_A \lambda_{A'} \psi_L
+{\rm \{L \to R\} }
,
\label{e:4pt}
\end{align}
where we have defined coefficients
$
G_\chi 
=
\frac{2(N_c^2 - 1)}{N_c^2}
$
and
$
G_\Delta
=
\frac{N_c + 1}{2 N_c}
$
.

\subsection{Bosonization}

We apply the Hubbard-Stratonovich transformation formula, 
$
e^{\beta_1 \beta_2/4 \alpha}
=
\int dz e^{-\alpha |z|^2 + \beta_1 z + \beta_2 z^*}
$, 
to the rearranged four-point interaction (\ref{e:4pt}).
For simplicity, we make further, but moderate assumptions 
in the formation of the chiral and diquark condensates. 
For chiral condensates, we assume that only the 
flavor-singlet condensates are formed, 
and for diquark condensates, that 
only the color-flavor-locked condensates are formed, 
i.e., 
$
s_{AA'}
=
\left<
\psi_L^{c \dagger} \tau_A \lambda_{A'} \psi_L
\right>
\propto 
\delta_{AA'}
$.
These assumptions allow us to 
bosonize the fermion vertex (\ref{e:fourpoint}) as
\begin{align}
&\exp \left[ -\frac{1}{8 N \Sigma^2} (J_{a\nu}^{ij})^2\right]
\nonumber
\\
=&
\int 
[d \phi][d \Delta]
\exp \left(
-
\frac{N \Sigma^2}{2 G_\chi}
2|\phi_f|^2
-
\frac{N \Sigma^2}{2 G_\Delta}
(
|\Delta_A^L|^2
+
|\Delta_A^R|^2
)
\right)
\nonumber
\\
&\qquad \times
\exp 
\left[
-
\left(
\phi_f^*
\psi_{R}^{f \dagger} \psi_{L}^f 
+
\psi_{L}^{f \dagger} \psi_{R}^f 
\phi_f
\right)
-
\frac{1}{2}
\left(
\Delta^{L*}_A
\psi_L^{c \dagger}  \tau_A \lambda_{A} \psi_L
+
\psi_L^\dagger     \tau_A \lambda_{A} \psi_L^c
\Delta^L_A
\right)
-
\{ L \to R\}
\right]
\nonumber
\\
=&
\int 
[d \phi][d \Delta]
\exp \left(
-
\frac{N \Sigma^2}{2 G_\chi}
2|\phi_f|^2
-
\frac{N \Sigma^2}{2 G_\Delta}
(
|\Delta_A^L|^2
+
|\Delta_A^R|^2
)
\right)
\exp
\left[
-
\Psi_L^\dagger 
S
\Psi_R
-
\Psi_R^\dagger 
S^\dagger
\Psi_L
\right]
,
\end{align}
where the measure 
$
[d\phi][d\Delta]
=
\prod_{f=u,d,s} \prod_{A=2,5,7}
d \phi_f
d \Delta_A^L d \Delta_A^R
$.
We have defined 
the Nambu-Gorkov spinors by
\begin{align}
\Psi
=
\frac{1}{\sqrt{2}}
\left(
\begin{matrix}
\Psi_R\\
\Psi_L
\end{matrix}
\right)
=
\frac{1}{\sqrt{2}}
\left(
\begin{matrix}
\psi_R   \\
\psi_L^c \\ 
\psi_L   \\
\psi_R^c
\end{matrix}
\right)
\end{align} 
and 
\begin{align}
\bar \Psi
=
\frac{1}{\sqrt{2}}
(\Psi_L^\dagger, \Psi_R^\dagger)
=
\frac{1}{\sqrt{2}}
(\psi_L^\dagger, \psi_R^{c \dagger}, \psi_R^\dagger, \psi_L^{c \dagger})
,
\end{align} 
and the $2\times N_C \times N_f (=18)$-dimensional 
order parameter matrix $S$ by
\begin{align}
S
=
\left(
\begin{matrix}
  \hat \phi  \mathbf{1}_{N_c}  & \Delta_A^L \tau_A \lambda_A \\
  \Delta_A^{R*} \tau_A \lambda_A   & \phi \mathbf{1}_{N_c}
\end{matrix}
\right)
,
\end{align}
where $\hat \phi = {\rm diag} (\phi_u, \phi_d, \phi_s)$ 
is a matrix in the flavor space.

By representing the mass and matter effect terms 
in the Nambu-Gorkov basis, 
we obtain the partition function as
\begin{align}
Z
=
\int 
[d \psi^\dagger] [d \psi]
[d \phi] [d \Delta]
\exp \left(
-
\frac{N \Sigma^2}{2 G_\chi}
2|\phi_f|^2
-
\frac{N \Sigma^2}{2 G_\Delta}
(
|\Delta_A^L|^2
+
|\Delta_A^R|^2
)
\right)
\exp \left[
-\bar \Psi 
\left(
\begin{matrix}
S+\m       & \tilde C \\
\tilde C  & S^\dagger + \m^\dagger
\end{matrix}
\right)
\Psi
\right]
,
\end{align}
where 
$\tilde C={\rm diag}(\omega, -\omega)$ 
and $\m$ is the extended mass matrix
\begin{align}
\m
=
\left(
\begin{matrix}
\hat m  \mathbf{1}_{N_c} & -\eta_A \tau_A \lambda_A \\
\eta_A^* \tau_A \lambda_A       & \hat m  \mathbf{1}_{N_c} 
\end{matrix}
\right)
\end{align}
with mass matrix $\hat m = {\rm diag}(m_u, m_d, m_s)$ 
in the flavor space.
We have introduced the external field $\eta_A$, 
which should be zero in the end of the calculation.
It is useful to clarify the meaning of the order parameters. 

Finally, the evaluation of the fermion integral is straightforward, 
which yields 
\begin{align}
Z
&=
\int 
[d \phi] [d \Delta]
\exp \left(
-
\frac{N \Sigma^2}{2 G_\chi}
2|\phi_f|^2
-
\frac{N \Sigma^2}{2 G_\Delta}
(
|\Delta_A^L|^2
+
|\Delta_A^R|^2
)
\right)
\left[
\det\;\hspace{-1.6mm}^{N/2} 
\left(
\begin{matrix}
S+\m       & \tilde z \\
\tilde z  & S^\dagger + \m^\dagger
\end{matrix}
\right)
\det\;\hspace{-1.6mm}^{N/2} 
\left(
\begin{matrix}
S+\m       & \tilde z^* \\
\tilde z^*  & S^\dagger + \m^\dagger
\end{matrix}
\right)
\right]^{1/2}
\nonumber
\\
&=
\int [d\phi] [d\Delta] e^{-2 N N_f N_c \Omega (\phi,\Delta; m, T, \mu)}
,
\end{align}
where $\tilde z = {\rm diag}(z, -z)$ with 
$z\equiv (\mu + i T) \mathbf{1}_{N_f} \mathbf{1}_{N_C}$. 
The square root over the determinant is given
because the number of the fermion measures 
is a half of that of the Nambu-Gorkov basis.
We have defined the effective potential $\Omega$ as the function
of the order parameters. 
In the thermodynamic limit, $N \to \infty$, 
the ground state of the model is determined by 
the set of the order parameters 
which minimizes the effective potential.

For the ground state solutions, 
we make a few Ansatzs for the order parameters. 
First, we set $\phi_f$ to be real, $\phi_f^* = \phi_f$. 
Second, we also set $\Delta_A^L$ and $\Delta_A^R$ 
to be real and 
$-\Delta_A^L=\Delta_A^R\equiv \Delta_A$.
Both assumptions are consistent for the formation of 
the scalar and parity-positive condensates in the ground state, 
which are favored by 
the finite quark mass term and the real $\eta_A$. 
Then, the effective potential is a function of six order parameters, 
$\phi_f$ with $f=u, d$, and $s$ and $\Delta_A$ with $A=2, 5$, and $7$.
Using these assumptions, 
the effective potential $\Omega$ becomes
\begin{align}
\Omega
=
&
\frac{B}{3}   \phi_f^2
+
\frac{A}{3} \Delta_A^2
\nonumber
\\
&-
\frac{1}{8 N_c N_f}
\left[
\ln \det (S        + \m + z)
+
\ln \det (S^\dagger + \m^\dagger - z)
+
\ln \det (S        + \m + z^*)
+
\ln \det (S^\dagger + \m^\dagger - z^*)
\right]
,
\label{e:pot}
\end{align}
where $A=3 \Sigma^2/(2N_c N_f G_\Delta)$ 
and $B=3 \Sigma^2/(2N_c N_f G_\chi)$ are defined. 
We can obtain the ground state by solving 
the six gap equations, 
$\partial \Omega/ \partial \phi_f=0$ and
$\partial \Omega/ \partial \Delta_A=0$,
simultaneously.

\subsection{Remarks}
To relate the order parameters in the ChRM model 
with the expectation values of the
fermion bilinears in the microscopic theory,
we use the external field derivatives as
\begin{align}
\left< \bar \psi_f \psi_f \right>
\equiv
-
\frac{1}{2N N_c N_f}
\frac{\partial \ln Z}{\partial m_f}
&=
\frac{2B}{3}
\phi_f
,
\\
s_{AA}
=
\left< \bar \psi^c \tau_A \lambda_A \gamma_5 \psi \right>
\equiv
-
\frac{1}{2N N_c N_f}
\frac{\partial \ln Z}{\partial \eta_A}
&=
\frac{2A}{3}\Delta_A
.
\end{align}
The order parameters $\phi_f$ and $\Delta_A$ are proportional 
to the chiral and the diquark condensates respectively, 
and then we simply use the values of $\phi_f$ and $\Delta_A$ 
to distinct each phases.

Note that, in the partition function,
the parameter $\Sigma$ can be absorbed 
by rescaling the order parameters, as well as 
the parameters, $T, \mu, m_f$, and $\eta_A$. 
Therefore, in the chiral limit $m_f=0$ 
(together with $\eta_A=0$), 
a change of $\Sigma$ affects only on 
the scale of the phase diagram, 
and the global structure of the phase diagram is invariant.
%
In fact, 
the parameter that can change the 
structure of the phase diagram is $B/A$, 
which is independent of $\Sigma$.
In our treatment, 
the ratio is fixed by the Fierz coefficients, 
and is obtained as $B/A = 3/8$ with $N_c=3$.
In Ref.~\cite{Vanderheyden:1999xp},
various structures of the phase diagrams 
has been found with $B/A$ changed.





\section{Three equal-mass flavors}
We first examine the ground state 
in the case with the exact flavor SU$(3)$ symmetry. 
For this purpose, we set $m_u=m_d=m_s=m$. 
Assuming that the flavor symmetry 
is not broken spontaneously, we can set 
the order parameters as 
$\phi_u=\phi_d=\phi_s\equiv \phi$ and
$\Delta_2=\Delta_5=\Delta_7\equiv \Delta$.

%
The effective potential is simplified to
the function of the two order parameters,
\begin{align}
\Omega
=
A \Delta^2
+
B \phi^2
-
\frac{1}{72}
\sum_\pm
\ln [(\sigma \pm z)^2 + \Delta^2]^8 [(\sigma \pm z)^2 + (2 \Delta)^2]
+
{\rm c.c. }
,
\label{e:cfl}
\end{align}
where $\sigma = \phi +m$ and we have set $\eta_A=0$.
Combining two gap equations, 
$\partial \Omega/\partial \phi=0$ and $\partial \Omega/\partial \Delta=0$, 
we can determine a ground state solution for given $T, \mu$, and $m$. 

It is easy to find that $\Delta=0$ is always a solution of 
the gap equation, since $\Delta$ appears as $\Delta^2$ in the effective potential.
Moreover, if $m=0$, $\phi=0$ is also a trivial solution for any $T$ and $\mu$. 
Then, in the chiral limit, 
there are generally four types of the solutions, 
(i)  $\phi = 0, \Delta = 0$, 
(ii) $\phi \neq 0, \Delta = 0$, 
(iii) $\phi = 0, \Delta \neq 0$, and
(iv) $\phi \neq 0, \Delta \neq 0$.
When $m\neq 0$, the solution $\phi =0$ no longer exists and 
is replaced by a small value proportional to $m$.

Let us first consider solutions with $\Delta=0$. 
When $\Delta=0$, the effective potential (\ref{e:cfl}) becomes identical 
with that analyzed in Ref.~\cite{Halasz:1998qr},
\begin{align}
\Omega 
=
\frac{G^2}{2} \phi^2 
-
\frac{1}{4}
\ln (\sigma^2 - z^2)
-
\frac{1}{4}
\ln (\sigma^2 - z^{*2})
,
\label{e:hs}
\end{align}
where $G^2 = 2B$ is defined. 
Therefore, the phase diagram described by the effective potential 
(\ref{e:cfl}) at $\Delta=0$ is the same as in Ref.~\cite{Halasz:1998qr}.
Several points on the phase structure 
with the effective potential (\ref{e:hs})
are summarized:
When $m=0$, we find a second-order phase boundary 
in the large $T$ and small $\mu$ region, and 
a first-order in the small $T$ and large $\mu$ region.
Two lines are connected at the TCP,
$(T_3, \mu_3 )
=
(
\frac{1}{2}\sqrt{\sqrt{2} + 1}G^{-1},
\frac{1}{2}\sqrt{\sqrt{2} - 1}G^{-1}
)
=
(0.776 G^{-1}, 0.322G^{-1})
$.
The transition temperature at $\mu=0$ is 
obtained as $T_0 = G^{-1}$, while the 
transition chemical potential at $T=0$ is
$\mu_0 = 0.528 G^{-1}$.
We use these two values of $T_0$ and $\mu_0$ 
for a normalization of $T$ and $\mu$ 
in the presentation of the phase diagram to remove 
the $\Sigma$ dependence as possible. 
If finite $m$ is introduced, 
the second-order phase transition line 
becomes smooth crossover, 
while the first-order line remains robustly.
The TCP also becomes the critical point.

We next consider the solutions with $\phi=0$. 
By setting $\phi=0$ and $m=0$, the effective potential becomes
\begin{align}
\Omega 
=
A \Delta^2 
-
\frac{1}{4}
\left[
\frac{8}{9}
\ln (  \Delta^2 + z^2)
+
\frac{1}{9}
\ln (4 \Delta^2 + z^2)
\right]
+
{\rm c.c.}
\label{e:phi0}
\end{align}
The gap equation for a nontrivial solution 
$\Delta\neq 0$ is obtained as
\begin{align}
A
-
\frac{1}{4}
\left[
\frac{8}{9}
\frac{1}{  \Delta^2 + z^2}
+
\frac{1}{9}
\frac{4}{4 \Delta^2 + z^2}
\right]
+
{\rm c.c.}
=
0
.
\label{e:phi0_nt}
\end{align}
For a large $T$ and/or $\mu$, this equation does not have 
a real solution of $\Delta$, 
which indicate that at some values of $T$ and $\mu$, 
its solution coalesces to the trivial solution $\Delta=0$, 
where the system reaches 
a second-order phase transition. 
It is easy to find a curve of the phase transition 
by setting $\Delta=0$ in Eq.~(\ref{e:phi0_nt})
\begin{align}
\frac{\mu^2 - T^2}{(\mu^2 + T^2)^2}
=
\frac{2A}{3}
\label{e:second}
.
\end{align}
The true phase structure should be, of course, 
determined by comparing the effective potential 
for all solutions of the gap equations, 
and the second-order phase transition line (\ref{e:second}) 
may be replaced by other phase structures.

\begin{figure}[tb]
\begin{center}
\includegraphics[width=0.4\textwidth]{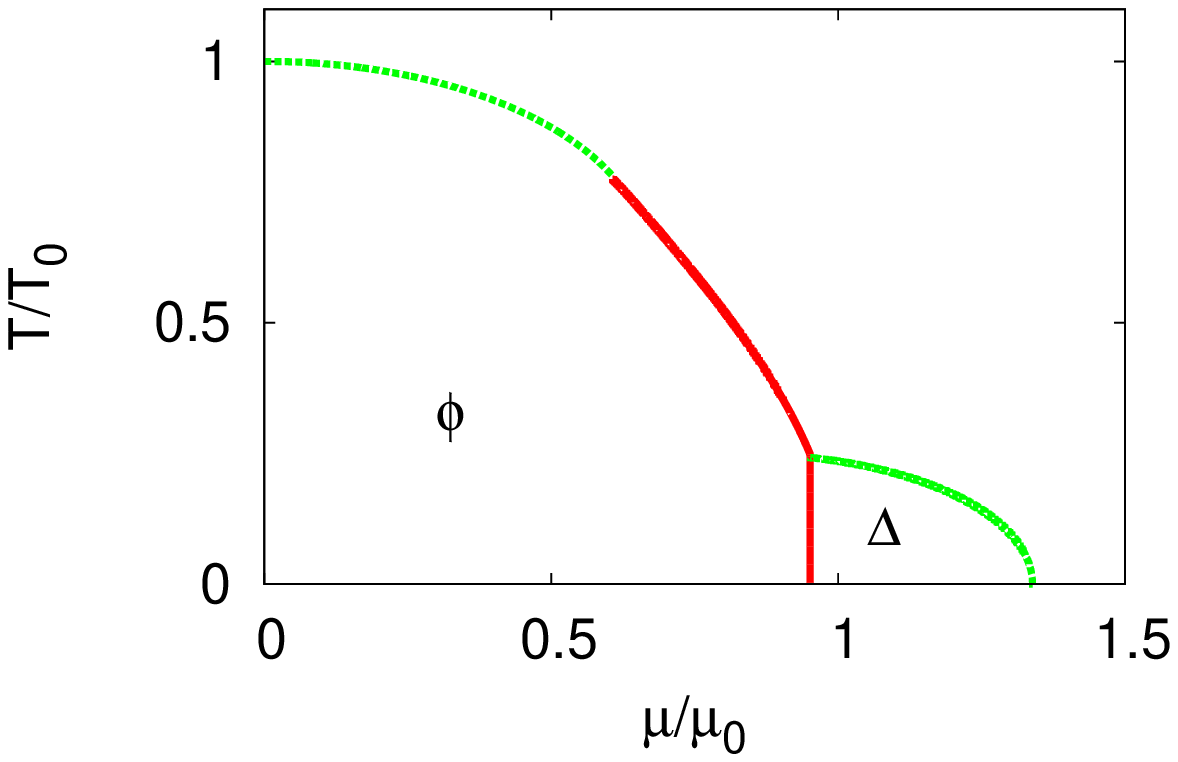}
\hfil
\includegraphics[width=0.4\textwidth]{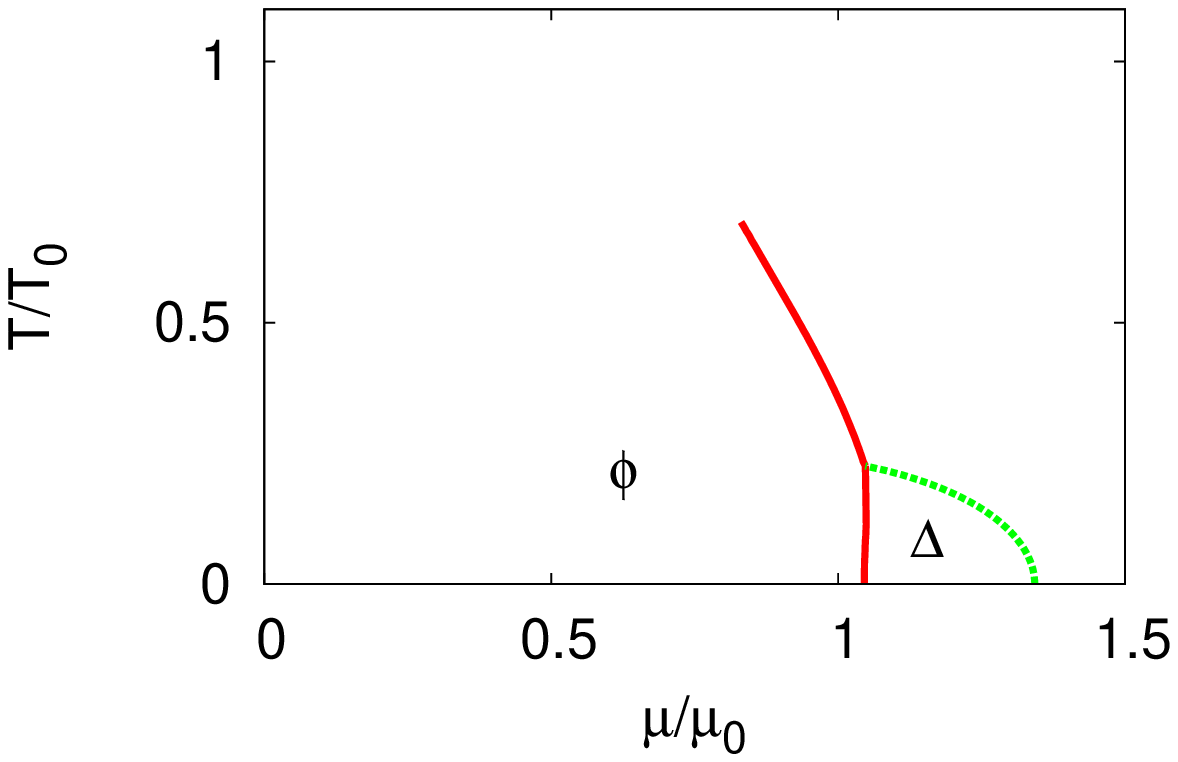}
\caption{
the phase diagrams with three equal-mass flavors. 
The left (right) panel shows the case with $m=0$ ($\Sigma m=0.1$).
The solid lines denote the first-order phase transitions, 
and the dotted lines the second-order phase transitions.
The largest order parameters in each phases are shown by its letter.
$T_0$ ($\mu_0$) is defined as the chiral phase transition 
temperature (chemical potential) on the $T$ ($\mu$) axis 
in the chiral limit when $\Delta=0$.
}
\label{f:cfl}
\end{center}
\end{figure}

To investigate the whole phase diagram, 
we have to numerically compare the effective potentials for 
all possible solutions.
The result is presented in Fig.~\ref{f:cfl} 
for the case with $m=0$ (left panel), 
and with $m\neq 0$ (right panel).

In the chiral limit, 
we find a $\Delta = 0$ solution becomes 
the ground state for 
a small chemical potential $\mu/\mu_0 \simlt 0.95$. 
In this region, the phase diagram is the same 
as that investigated in Ref.~\cite{Halasz:1998qr}:
There is the second-order phase transition line 
between the $T$ axis and the TCP,
and the first-order line from TCP to the large $\mu$ region.

When a chemical potential exceeds a critical value, 
however, we find a first-order phase transition to 
the CFL phase at a small temperature. 
In this phase, the chiral order parameter $\phi$ becomes zero.
We stress that the phase transition 
to the CFL phase in the ChRM model is remarkable 
because in the construction of the ChRM model,
the ratio of the couplings between
the quark-antiquark and the quark-quark channels
is given by the symmetry consistent with QCD, 
not tuned so that the phase transition can be reproduced.

Unfortunately, on the other hand, 
the CFL condensate $\Delta$ continuously goes to zero 
not only as $T$ is increased, but also as $\mu$ is. 
In QCD,
the second-order phase transition at a large $\mu$ 
is not expected, since the Cooper instability 
remains 
even if the attractive interaction is infinitely small. 
This unphysical phase transition may be explained by 
the absence of the Fermi surface in the ChRM model,
which is regarded as a model without the spacial dimension.
We consider that, at such a region,
the model reaches a limitation.
%
Similar structures are found in the ChRM models 
for two-color QCD with 
the diquark baryon condensates\cite{Vanderheyden:2001gx,Klein:2004hv}, 
as well as in the study on the 2SC phase 
in the ChRM model\cite{Vanderheyden:1999xp}.

When $m\neq0$ (the right panel of Fig.~\ref{f:cfl}), 
we find qualitative and quantitative changes from the 
case with $m=0$. 
Due to the nonzero $m$, $\phi$ has a small value 
even in the symmetric and the CFL phases. 
The second-order chiral phase transition line 
is washed out to become a crossover, 
and then the TCP becomes a critical point. 
Note that the second-order phase transition line for $\Delta$ 
remains. 
The first-order chiral phase transition line is
pulled up in the larger $T$ and $\mu$ directions. 
The phase transition line between 
the chirally-broken phase and the CFL phase also 
shifts to the larger $\mu$ region. 
Although the second-order CFL phase transition line
moves to extend the CFL phase, 
the CFL phase shrinks in total.

\section{2+1 flavors}
We next concentrate on the response of the model to 
the asymmetry between the light up and down (ud) quark flavors and 
the mid-light strange quark flavor. 
In order to see this effect,
we set the quark masses as $m_u=m_d=m$ and $m_s \neq m$. 
We assume that 
the flavor symmetry is not broken further spontaneously, 
and set the order parameters as 
$\phi_u=\phi_d=\phi$ and $\Delta_5=\Delta_7=\Delta_s$. 
We also write $\Delta_2 = \Delta$ for convenience.
Due to the asymmetry between the ud and strange quarks,
the typical symmetry in the CFL phase 
SU$_{c+L+R}(3)$ is not realized.
Nevertheless, we call the phase 
with $\Delta \neq 0$ and $\Delta_s\neq 0$ the CFL phase.
Also, the phase with $\Delta \neq 0$ and $\Delta_s = 0$ 
is defined as the 2SC phase

Under the parametrization, 
the effective potential becomes the function of the 
four order parameters as
\begin{align}
\Omega
=
&
\frac{A}{3}( \Delta^2 + 2 \Delta_s)
+
\frac{B}{3}( 2\phi^2 + \phi_s)
\nonumber
\\
&-
\frac{1}{72}
\sum_{\pm} \ln 
\left[
\{ (\sigma \pm z)^2 + \Delta^2 \}^3
\{ (\sigma \pm z)(\sigma_s \pm z) + \Delta_s^2 \}^4
\{ (\sigma_s \pm z)^2((\sigma \pm z)^2 +\Delta^2) 
  + 4 \Delta_s^2((\sigma \pm z)(\sigma_s \pm z)+ \Delta_s^2 ) \}
\right]
\nonumber
\\
&+ {\rm c.c. }
,
\label{e:2sc}
\end{align}
where $\sigma = \phi + m$, $\sigma_s = \phi_s + m_s$,
and we set $\eta_A=0$. 
For the general case with $\phi\neq \phi_s$
and $\Delta\neq \Delta_s$, 
the determinant part under the logarithm
becomes complicated.
By setting $m_s=m$, $\phi_s=\phi$ and $\Delta_s =\Delta$, 
we recover the effective potential in the three equal-mass 
limit (\ref{e:cfl}). 
Another interesting limit is the 2SC phase, 
where $\Delta_s=0$, 
in which the effective potential is separated to 
the ud quark sector and the strange quark sector as
\begin{align}
\Omega 
= 
\Omega_{ud}(\phi, \Delta) + \Omega_s (\phi_s)
,
\end{align} 
where
\begin{align}
\Omega_{ud}(\phi, \Delta)
=
\frac{A}{3} \Delta^2 
+ 
\frac{2B}{3}\phi^2
-
\frac{1}{72}\sum_{\pm}
\left[
4 \ln \left(
(\sigma \pm z)^2 +\Delta^2 
\right)
+
2 \ln \left(
\sigma \pm z
\right)^2
\right]
+{\rm c.c}
\label{e:pot_ud}
\end{align}
and
\begin{align}
\Omega_{s}(\phi_s)
=
\frac{B}{3}\phi_s^2
-
\frac{1}{72}
\sum_{\pm}
\ln \left(
\sigma_s \pm z
\right)^6
+{\rm c.c.}
\label{e:pot_s}
\end{align}
The effective potential for 
the strange quark flavor (\ref{e:pot_s}) is 
equivalent to the one of the conventional ChRM model
without diquark condensates\cite{Halasz:1998qr}, 
whose phase structure is summarized in sec.~IV.

The effective potential (\ref{e:pot_ud}) can be 
compared 
%
to the one in Ref.~\cite{Vanderheyden:1999xp},
where two light quark flavors are introduced
and the strange quark degree of freedom is neglected. 
We first point out that
the ratio of the coefficients of the quadratic terms 
of the order parameters, 
which can be read from (\ref{e:pot_ud}) as
$(2B/3)/(A/3)=2B/A = 3/4$ since $B/A=3/8$,
is equal to that appearing in the model
in Ref.~\cite{Vanderheyden:1999xp}.
The reproduction of $3/4$ is important since
the phase structure is sensitive to this ratio.

Interestingly, however,
these two effective potentials are not completely equivalent, 
and match only if $T=0$ or $\Delta=0$.
This is because we use the different temperature scheme,
or the matter effect matrix (\ref{e:mem}),
from that used in Ref.~\cite{Vanderheyden:1999xp}.
In the scheme used in Ref.~\cite{Vanderheyden:1999xp}, 
the effective temperature is introduced with opposite 
signs for two flavors%
\footnote{%
Superficially, it seems to break the flavor symmetry, 
but, indeed, the symmetry is hold if the chemical potentials 
for two flavors are equal\cite{Vanderheyden:2005ux}.
}%
.
On the other hand,
in the scheme we used here,
the temperature is introduced proportionally to the
identity in the flavor space.
The relation of these schemes is discussed 
in detail in Ref.~\cite{Vanderheyden:2005ux}.
In Ref.~\cite{Vanderheyden:2005ux},
it is found that there is a unitary matrix 
in the flavor and the Matsubara space,
which is a subspace of the zero-mode space,
to rotate the one scheme to the other.
The fermion bilinears corresponding to the chiral condensates 
are invariant under the rotation, 
but the diquark condensates are not.
This is the formal reason why
we find that the two effective potentials
are equivalent not only at $T=0$, 
but also at $\Delta =0$.

By considering the microscopic theory,
a reason of 
the choice of the flavor-antisymmetric scheme can be explained
as follows\cite{Vanderheyden:2005ux}:
To make the flavor-antisymmetric quark pair condensate be 
independent of the imaginary time $\tau$,
only the terms of two quark fields with 
the opposite signs of the Matsubara frequencies should be nonzero
in the Fourier summation over the Matsubara frequencies.
This indicates that the temperature term, 
or the lowest Matsubara frequencies, 
should have the opposite signs for different flavors,
because, in the diquark pairing, two quarks have different flavors
in our treatment.


%


Nevertheless, we use the flavor-symmetric scheme in this paper
because of the following reasons.
First, if we construct the ChRM model relying only on the symmetry,
we can not find a principle which scheme has to be adopted.
Both treatments are consistent with the QCD symmetry, 
the anti-hermiticity at $\mu=0$, and the chiral symmetry for all $T$ and $\mu$
of the Dirac operator.
Second, 
because we treat the ChRM model in the case with three flavors,
in order to adopt the flavor-antisymmetric scheme,
we have to concern contributions
from three combinations of the three flavors.
This might make the effective potential more complicated,
and we try to make the model as simple as possible.
Indeed, the resulting phase diagram,
which will be shown below,
is qualitatively equivalent to 
the model with the flavor-antisymmetric scheme at $\Delta_s=0$.
In other words,
they have the same global structure, 
or topology, of the phases.
It suggests that the qualitative structure of the phase diagram is not 
sensitive to the selection of the temperature dependence schemes,
as long as they hold the symmetry.

\begin{figure}[tb]
\begin{center}
\includegraphics[width=0.4\textwidth]{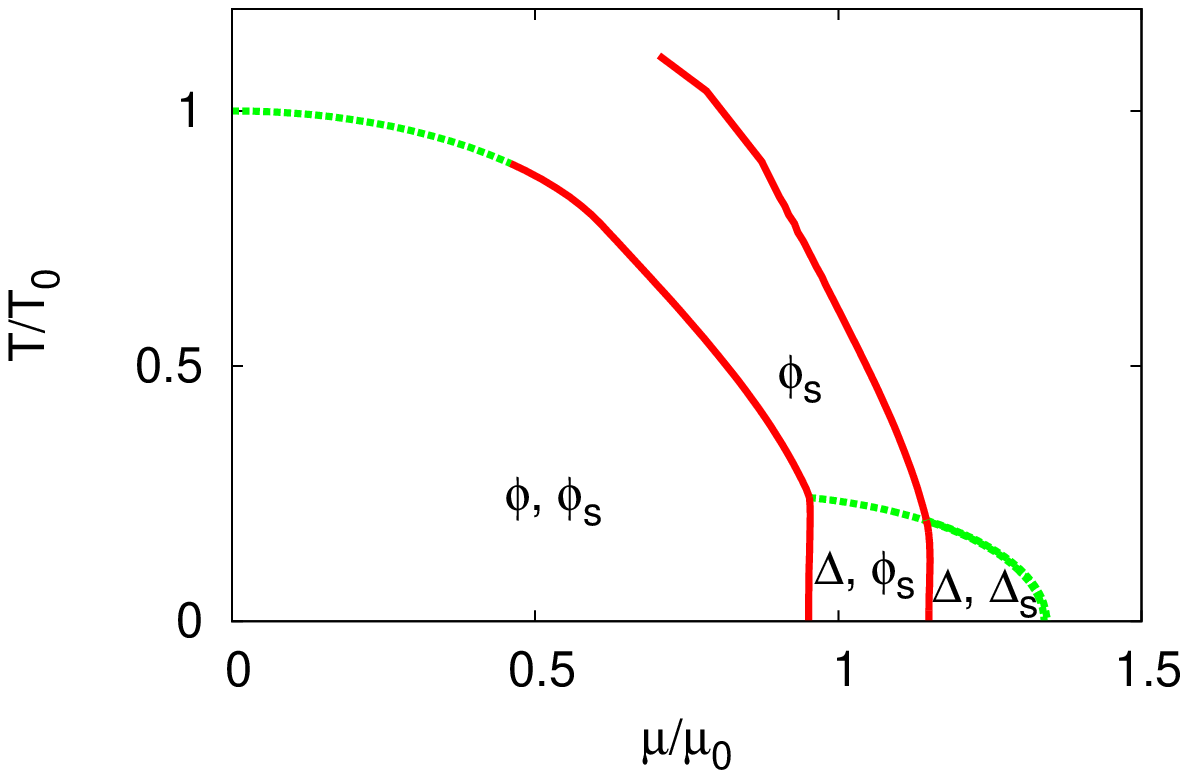}
\hfil
\includegraphics[width=0.4\textwidth]{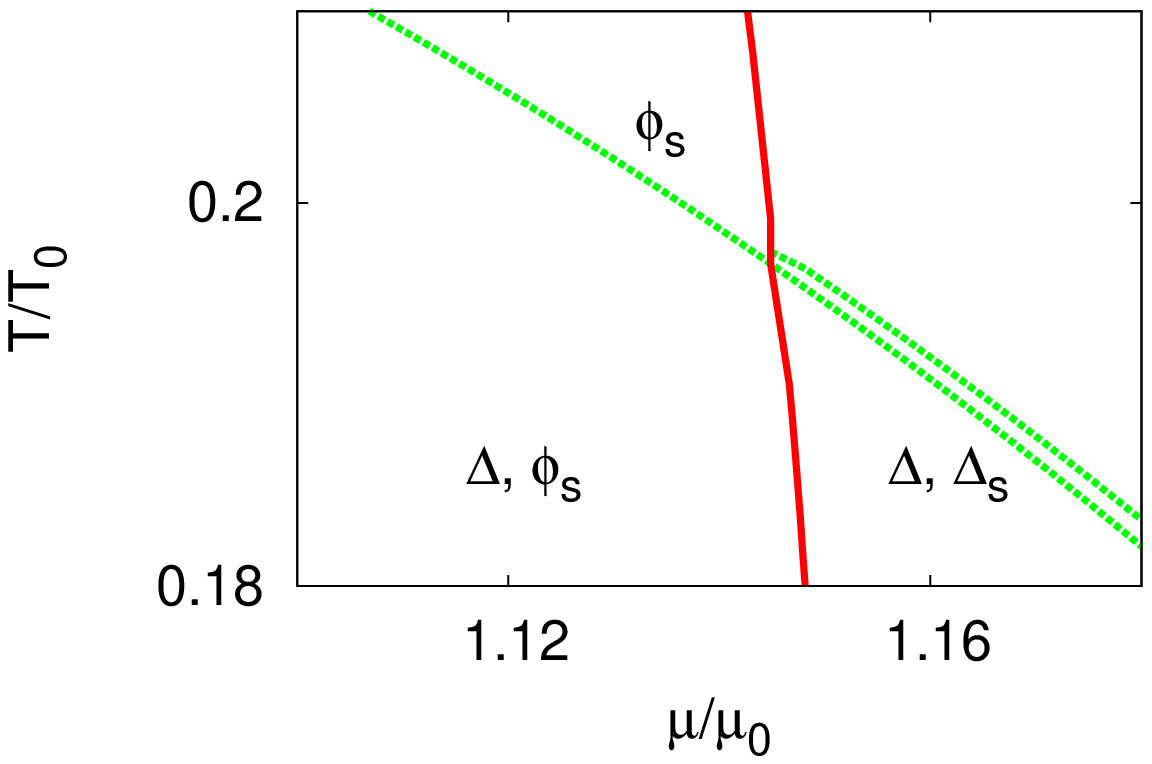}
\caption{
The phase diagram with two massless and one massive quark flavors,
$m_u=m_d=0$ and $\Sigma m_s = 0.14$.
Notations are the same with Fig.~1.
In the narrow slit between the symmetric phase and
the CFL phase, 
which can be seen in the right panel,
only $\Delta_s$ has a large value.
}
\label{f:2sc}
\end{center}
\end{figure}

We next present the resulting phase diagram.
Similar to the case with three equal-mass flavors, 
we need numerical calculations to evaluate the whole phase diagram. 
The result with $m=0$ and $m_s\neq 0$ is shown in Fig.~\ref{f:2sc}.
At a small chemical potential, 
the diquark condensates do not have finite values in the ground state, 
and the phase diagram is the same with the model 
without the diquark condensates. 
Because for the ud quark flavors, 
there are no symmetry-breaking mass terms,
the second-order phase transition line, as well as the TCP, exist. 
On the contrary, for the strange quark flavor, 
finite $m_s$ breaks the chiral symmetry explicitly, 
and then the second-order phase transition line becomes 
crossover. The TCP also becomes a critical point. 
In the symmetric phase, $\phi, \Delta$ and $\Delta_s$ 
become exactly zero, but $\phi_s$ never becomes zero 
due to the finite symmetry breaking term $m_s$.

As $\mu$ is increased with $T$ kept small, 
a first-order phase transition to the 2SC phase is first observed. 
In the 2SC phase, not only $\Delta$, 
but also $\phi_s$ have large values. 
If $\mu$ is further increased, 
we next observe a phase transition to the CFL phase, 
where both $\Delta$ and $\Delta_s$ have large values. 
Note that because the flavor symmetry is explicitly broken, 
generally $\Delta \neq \Delta_s$ in the CFL phase. 
The diquark condensates become zero continuously 
when $T$ and $\mu$ are increased.  
As seen in the right panel of Fig.~\ref{f:2sc},
there is a narrow slit between two second-order phase transition lines 
for $\Delta$ and $\Delta_s$.
In this area, only $\Delta_s$ has a large value. 
Four phase transition lines, 
two horizontal second-order phase transition lines for $\Delta$,
and the vertical first-order phase transition lines 
for $\phi_s$ and $\Delta_s$, 
meet at one point. 
The second-order phase transition line for $\Delta_s$ 
touches to the first-order phase transition line for $\phi_s$.

We consider the sequence of 
the melting of the diquark condensates. 
In the CFL phase, 
as $T$ is increased with $\mu$ fixed, 
we find that the $\Delta$ becomes zero firstly, 
and the $\Delta_s$ secondary. 
In the analysis of 
the NJL models\cite{Buballa:2001gj, Fukushima:2004zq}, 
on the contrary, 
it was found that $\Delta_s$ melts firstly and 
$\Delta$ secondary, and then 
the phase with only $\Delta_s$ finite is not found. 
The 2SC phase (where only $\Delta$ finite) above the CFL phase 
may also be expected in QCD. 
The reason is explained 
as follows\cite{Fukushima:2004zq,Fukushima:2010bq}:
Around the melting temperature,
the quark Fermi spheres are smeared and then
the sizes of gaps are mainly dominated by the density of states.
Because the larger $m_s$ makes the strange quark Fermi sphere smaller, 
its density of states is smaller than that of the ud quarks,
which results in the smaller $\Delta_s$ than $\Delta$.
This indicates the preceding melting of $\Delta_s$.
We consider that the contradicting result of our model is
due to the absence of the Fermi sphere in our treatment, 
and, unfortunately, 
that this is a limitation of our model.

\section{Summary and Discussion}

We have studied a ChRM model which can treat
the competition between 
the chiral and the diquark condensates 
in the case with three flavors, 
to investigate the phase diagram with the CFL phase.
In order to describe the color superconductivity, 
we have introduced the color and Lorentz indices 
to the truncated Dirac matrix. 
The random matrices mimicking the gauge fields 
are taken to be real matrices, whose elements 
are distributed according to the Gaussian weight. 
After the integration over the random matrices, 
we are left with the four fermion interaction term, 
which contains not only quark-antiquark interaction vertices, 
but also quark-quark vertices. 
Using Fierz transformations,
the ratio of the coefficients is uniquely determined. 
Applying the bosonization techniques, 
we finally derive the effective potential as the function 
of the order parameters, 
the chiral condensates and the diquark condensates. 

The phase diagram on the $T$-$\mu$ plane is 
calculated by solving the gap equations simultaneously. 
In the case with three equal-mass quark flavors, 
we find the CFL phase in the large chemical potential region, 
while the chirally-broken phase in the small chemical potential region. 
In the region where the CFL condensate is zero, 
the phase diagram is equivalent to that obtained 
from the conventional ChRM model without the diquark condensations.
The CFL condensates become zero continuously, 
as $T$ and/or $\mu$ are increased.
The unphysical phase transition at large $\mu$ may be
considered as the result of the fact that 
the ChRM model does not contain the Fermi surface.
When finite quark mass $m$ is introduced, 
the phase transition line for the CFL phase 
moves towards the larger chemical potential region.
We also find that the region of the CFL phase becomes smaller
as $m$ is increased.

For the case with 2+1 flavors, $m_u=m_d=0$ and $m_s\neq 0$, 
we find both the CFL and the 2SC phases.
Moreover, we find the phase where 
$\Delta_s$ has a finite value and $\Delta=0$. 
Such phase is not found in the NJL model,
and not expected in QCD.
We also consider this phase may be an model artifact 
due to the absence of the Fermi surface, 
but can not be excluded from the viewpoint of the symmetry.


Although there are unphysical points, 
the ChRM model studied in this paper 
can consistently address both 
the chirally broken phase and the CFL phase.
In addition, with the finite asymmetry between 
the ud quarks and the strange quark, 
the model also can show the 2SC phase. 
Because the ChRM model includes the Dirac matrix, 
investigation of its eigenvalue distribution, the Dirac spectrum, 
may be possible.
It may be possible to understand 
the phase transition into the diquark-condensed states 
in the context of the moving of the Dirac eigenvalues. 
Furthermore, in the microscopic region, 
there might be the universal structure of the Dirac spectrum, 
which can be compared to the other models 
at high density\cite{Akemann:2007rf,Yamamoto:2009ey}. 
The applications in these directions
are postponed to future studies.

Finally, we consider an outlook or a possible extension of the model. 
One of the most important effect we neglect
is the U(1) breaking axial anomaly effect. 
It is known that, in the ChRM model, 
the anomaly effect introduces the flavor mixing, which
changes the phase diagram drastically\cite{Sano:2009wd,Fujii:2009fm,Fujii:2010xy}. 
Indeed, we find that the effective potential 
in the case with 2+1 flavors and $\Delta_s=0$ 
can be separated to the ud quark sector 
and the strange quark sector, 
as a result of the absence of the mixing effect. 
Moreover, on the phase diagram around mid chemical potential, 
it is suggested that the anomaly effect, 
which also mixes the chiral and the diquark order parameters,  
plays a crucial role on the phase structure 
with the various superconducting 
phases\cite{Hatsuda:2006ps,Abuki:2010jq,Basler:2010xy}.
We are then interested to combine 
the treatment of the ChRM model 
with the color superconducting phases presented here,
and of the ChRM model with the axial anomaly in Ref.~\cite{Sano:2009wd}
to investigate the effects of the axial anomaly 
on the phase structure with the color superconducting phases. 
It will allow us to discuss the phase structure 
at finite temperature and density
from a viewpoint of the symmetry.



\acknowledgements
The authors are grateful to 
members of Komaba nuclear and particle theory group for their
interests in this work and encouragements. 
They especially thank Tetsuo Matsui and 
Hirotsugu Fujii for useful comments.
TS is supported by 
the Japan Society for the Promotion 
of Science for Young Scientists.


\begin{thebibliography}{99}
  


\bibitem{CMP_QCD}
  K.~Rajagopal and F.~Wilczek,
  arXiv:hep-ph/0011333.


\bibitem{Fukushima:2010bq}
  K.~Fukushima, T.~Hatsuda,
  Rept.\ Prog.\ Phys.\  {\bf 74}, 014001 (2011).
  [arXiv:1005.4814 [hep-ph]].


\bibitem{Alford:2007xm}
For a recent review,  M.~G.~Alford, A.~Schmitt, K.~Rajagopal and T.~Schafer,
  Rev.\ Mod.\ Phys.\  {\bf 80}, 1455 (2008).


\bibitem{Hatsuda:1994pi}
  T.~Hatsuda, T.~Kunihiro,
  Phys.\ Rept.\  {\bf 247}, 221-367 (1994).
  [hep-ph/9401310].


\bibitem{Buballa:2003qv}
  M.~Buballa,
  Phys.\ Rept.\  {\bf 407}, 205-376 (2005).
  [hep-ph/0402234].


\bibitem{DeTar:2011nm}
  For a recent review,
  C.~DeTar,
  arXiv:1101.0208 [hep-lat].


\bibitem{deForcrand:2010ys}
  For a recent review,
  P.~de Forcrand,
  PoS {\bf LAT2009}, 010 (2009).
  [arXiv:1005.0539 [hep-lat]].

\bibitem{Barrois:1977xd}
  B.~C.~Barrois,
  Nucl.\ Phys.\  {\bf B129}, 390 (1977).


\bibitem{Bailin:1983bm}
  D.~Bailin, A.~Love,
  Phys.\ Rept.\  {\bf 107}, 325 (1984).





\bibitem{ChRM}
E.~V.~Shuryak and J.~J.~M.~Verbaarschot,
Nucl.\ Phys.\  A {\bf 560}, 306 (1993); 
%
  A.~D.~Jackson and J.~J.~M.~Verbaarschot,
  Phys.\ Rev.\  D {\bf 53}, 7223  (1996);
  T.~Wettig, A.~Sch\"afer and H.~A.~Weidenm\"uller,
  Phys.\ Lett.\  B {\bf 367}, 28 (1996) 
  [Erratum {\it ibid}.\  B {\bf 374} (1996) 362];
for review,  J.~J.~M.\ Verbaarschot and T.~Wettig,
Ann.\ Rev.\ Nucl.\ Part.\ Sci.  {\bf 50}, 343 (2000)
[arXiv:hep-ph/0003017].


\bibitem{Halasz:1998qr}
  A.~M.~Halasz, A.~D.~Jackson, R.~E.~Shrock, M.~A.~Stephanov, J.~J.~M.~Verbaarschot,
  Phys.\ Rev.\  {\bf D58}, 096007 (1998).




\bibitem{Asakawa:1989bq}
  M.~Asakawa, K.~Yazaki,
  Nucl.\ Phys.\  {\bf A504}, 668-684 (1989).

\bibitem{Barducci:1989wi}
  A.~Barducci, R.~Casalbuoni, S.~De Curtis, R.~Gatto, G.~Pettini,
  Phys.\ Lett.\  {\bf B231}, 463 (1989).

  
\bibitem{Vanderheyden:1999xp}
  B.~Vanderheyden, A.~D.~Jackson,
  Phys.\ Rev.\  {\bf D61}, 076004 (2000)
;  
  B.~Vanderheyden, A.~D.~Jackson,
  Phys.\ Rev.\  {\bf D62}, 094010 (2000).

\bibitem{Banks:1979yr}
  T.~Banks, A.~Casher,
  Nucl.\ Phys.\  {\bf B169}, 103 (1980).



\bibitem{Fukushima:2008su}
  K.~Fukushima,
  JHEP {\bf 0807}, 083 (2008).





\bibitem{Vanderheyden:2001gx}
  B.~Vanderheyden, A.~D.~Jackson,
  Phys.\ Rev.\  {\bf D64}, 074016 (2001).


\bibitem{Klein:2004hv}
  B.~Klein, D.~Toublan, J.~J.~M.~Verbaarschot,
  Phys.\ Rev.\  {\bf D72}, 015007 (2005).


\bibitem{Sano:2009wd}
  T.~Sano, H.~Fujii, M.~Ohtani,
  Phys.\ Rev.\  {\bf D80}, 034007 (2009).


\bibitem{Janik:1996nw}
  R.~A.~Janik, M.~A.~Nowak, G.~Papp, I.~Zahed,
  Nucl.\ Phys.\  {\bf B498}, 313-330 (1997).







\bibitem{Vanderheyden:2005ux}
  B.~Vanderheyden, A.~D.~Jackson,
  Phys.\ Rev.\  {\bf D72}, 016003 (2005).


\bibitem{Buballa:2001gj}
  M.~Buballa, M.~Oertel,
  Nucl.\ Phys.\  {\bf A703}, 770-784 (2002).


\bibitem{Fukushima:2004zq}
  K.~Fukushima, C.~Kouvaris, K.~Rajagopal,
  Phys.\ Rev.\  {\bf D71}, 034002 (2005).






\bibitem{Yamamoto:2009ey}
  N.~Yamamoto, T.~Kanazawa,
  Phys.\ Rev.\ Lett.\  {\bf 103}, 032001 (2009).


\bibitem{Akemann:2007rf}
  G.~Akemann,
  Int.\ J.\ Mod.\ Phys.\  {\bf A22}, 1077-1122 (2007).
  [hep-th/0701175].


\bibitem{Fujii:2009fm}
  H.~Fujii, T.~Sano,
  Phys.\ Rev.\  {\bf D81}, 037502 (2010).


\bibitem{Fujii:2010xy}
  H.~Fujii, T.~Sano,
  Phys.\ Rev.\  {\bf D83}, 014005 (2011).

\bibitem{Hatsuda:2006ps}
  T.~Hatsuda, M.~Tachibana, N.~Yamamoto, G.~Baym,
  Phys.\ Rev.\ Lett.\  {\bf 97}, 122001 (2006).


\bibitem{Abuki:2010jq}
  H.~Abuki, G.~Baym, T.~Hatsuda, N.~Yamamoto,
  Phys.\ Rev.\  {\bf D81}, 125010 (2010).

\bibitem{Basler:2010xy}
  H.~Basler, M.~Buballa,
  Phys.\ Rev.\  {\bf D82}, 094004 (2010).





\end{thebibliography}
\end{document}